\documentclass[twocolumn,showpacs,preprintnumbers,amsmath,amssymb]{revtex4}

\usepackage{graphicx}
\usepackage{dcolumn}
\usepackage{bm}

\begin{document}

\title{Quasiparticle transitions in a charge-phase qubit probed by rf-oscillations}

 \author{J.~K\"{o}nemann\email{jens.koenemann@ptb.de}, H.~Zangerle, B.~Mackrodt, R.~Dolata, and A.\,B.~Zorin}
 \affiliation{Physikalisch-Technische Bundesanstalt, Bundesallee 100, 38116 Braunschweig, Germany}%

\date{\today}

\begin{abstract}
We  investigated  transitions in an Al charge-phase qubit of
SQUID-configuration which was inductively coupled to an rf tank
circuit that made it possible to read out the state by measuring
the Josephson inductance of the qubit. Depending on the flux and
charge bias and on the amplitude of the rf-oscillations, we probed
either the ground state or  a dynamic change of the qubit states
which we attributed to stochastic single quasiparticle tunneling
onto and off the qubit island, involving an exchange of energy
with the qubit. Within the scope of this model, a selection rule
for quasiparticle-induced transitions in the qubit is discussed.
\end{abstract}

\maketitle

Superconducting circuits  based on small Josephson junctions and
enabling the tunneling of single Cooper pairs offer great
opportunities for electronics and quantum computing.  The class of
the single Cooper pair devices comprises  Bloch transistors
\cite{Averin-Likharev,tuominen,chargeexp}, quantum electrometers
\cite{CP-R-electr-theo,CP-R-electr-exp-theo,CP-electr-HUT},
2$e$-pumps \cite{Geerligs}, etc. A prerequisite for the regular
operation of these devices is that no quasiparticle (QP)
transitions occur and that the so-called even-parity state of the
small superconducting island is thus maintained. If this
requirement is not fulfilled, even  a single QP can instantly
change the island charge, leading to a shifting of the operation
point.

There are several ways of observing the even-odd states of the
Bloch transistor island experimentally. These  include: (i) the
measuring of the gate dependence of the Josephson switching
current (see, e.g., Ref.\cite{chargeexp}), (ii) the monitoring of
the supercurrent peak
 \cite{tuominen,Amar}, and (iii) the
measuring of the island charge $q$ in the zero-current-biased
transistor (i.e. Cooper pair box) by means of a capacitively
coupled SET electrometer
\cite{Bouchiat,lehnert,lukens,echternach}. Recently, Naaman and
Aumentado \cite{aumentado} have investigated the parity states of
a current-biased Bloch transistor included in a resonance
$LC$-circuit driven by a 500~MHz harmonic signal. In this circuit,
the charge-dependent parameter was the Josephson inductance
$L_{\rm J}(q)$ of the Bloch transistor in the ground state.

 In this paper, we will address the problem of single QPs tunneling in Josephson qubit
 circuits. The operation of  charge qubits
\cite{qubit-NEC} is, by nature, sensitive to the incoherent
tunneling of unpaired electrons  because they change the charge
state of the system instantaneously and stochastically
\cite{qubit-Saclay, qubit-Chalmers}. This makes the setting of the
optimal working point of  the qubit difficult and increases the
 decoherence \cite{qubit-Chalmers,Zorin-JETP}.

 Our circuit comprises  a Bloch transistor which is included in a
superconducting ring
 inductively coupled to an rf-driven tank circuit \cite{zang,Zorin-JETP,cpqb2,imt2}. In
such an rf-SQUID configuration, the Bloch transistor is
galvanically decoupled from the measuring circuit, which in
general leads to a reduction in the density of non-equilibrium QPs
that are able to enter the island.  The effect of the QP
transitions manifested itself in excitations of this qubit and was
detected by measuring the change in the Josephson inductance value
$L_{\rm J}(q)$. We will demonstrate that in the general survey,
the dynamics of our circuit can
 be described in a similar way as in the model of Lutchyn \emph{et al.}
 \cite{glazman2}, which has been  published recently and addresses the problem of the energy relaxation in the charge qubit
 caused by a single QP.  In contrast to Ref.\cite{glazman2}, however, we will  focus in this paper on the reverse process, i.e. on the energy
 transfer
 from the QP  to the qubit, leading to an excitation of the qubit. Moreover, we will derive a selection rule
  showing that the rates of the QP transitions strongly depend on the initial qubit state.

\begin{figure}
\includegraphics[width=0.77\linewidth]{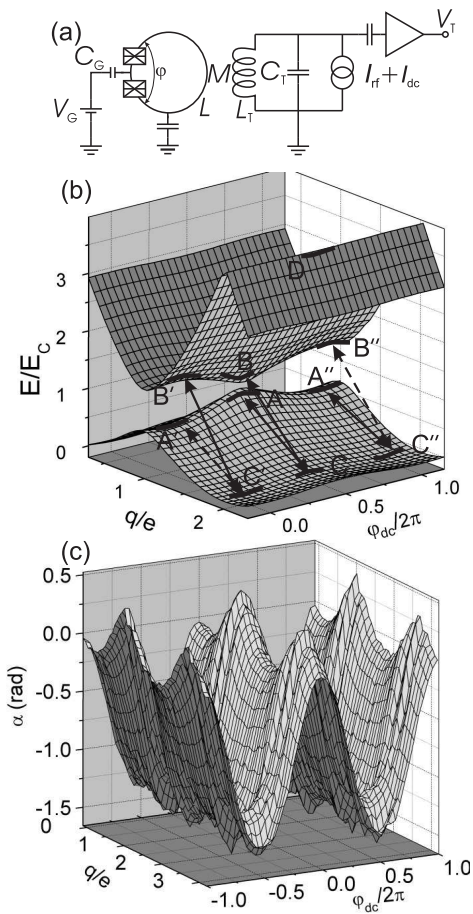}
\caption{\label{fig:fig1} (a) Diagram of the measurement set-up.
The core element is a double Josephson junction (the two crossed
boxes) with a capacitive gate coupled to its island, i.e. the
Bloch transistor, embedded in a macroscopic superconducting loop.
The loop is inductively coupled to an rf-driven tank circuit which
is capacitively coupled to a cold preamplifier.
 (b) Energy band diagram of the circuit, calculated for the parameters found in the experiment. The thick-line arcs show the variations of the phase $\varphi$ caused by oscillations in the tank
 circuit. The arrows denote the quasiparticle-induced transition
 between the qubit states (the dashed arrows indicate
  transitions suppressed due to the destructive interference effect,
 see text below).
  (c) Phase shift $\alpha$ in the
tank circuit, measured as a function of the dc-value of the
Josephson phase $\varphi_{\rm dc}$  and the island charge $q$
 for an amplitude
$a=0.56$ rad of rf-oscillations of the phase.}
\end{figure}
In our Al qubit circuit (see Fig.\,\ref{fig:fig1} (a)), the
superconducting loop which closes the Bloch transistor has a small
geometrical inductance $L\ll L_{\rm J}$. Thus, the total
inductance of the closed loop is determined mostly by the
transistor's Josephson inductance.
 The transistor is operated as a Cooper pair box (qubit) whose distinct
quantum states - which have the ground state energy $E_0$ and the
excited state energy $E_1$ (see Fig. 1(b)) - are associated with
different Bloch-bands of the system \cite{sqc}. The quantum states
of the transistor also
 depend on the phase coordinate $\varphi_{\rm dc}$ determined by the external
magnetic flux $\Phi_{\rm dc}$ that is applied to the  loop
$\varphi_{\rm dc}=2\pi\Phi_{\rm dc}/\Phi_0$, with $\Phi_0$ being
the flux quantum.  The Josephson inductance $L_{\rm J}$ is related
to the local curvature of the corresponding energy surface $E_{\rm
n}(q,\varphi)$, i.e. $1/L_{\rm J}(n,q,\varphi)\propto
\partial^2E_{\rm n}(q,\varphi)/\partial\varphi^2,n=0$ and $1$.
  Thus, the qubit eigenstates can be
identified by means of the Josephson inductance of the transistor
which is probed by small rf-oscillations in the loop, with the
resulting phase $\varphi(t)=\varphi_{\rm dc}+a\sin{(2\pi ft)}$ and
with $a$ being proportional to the amplitude of the
rf-oscillations in the tank circuit induced by an rf-driving
current. The black arcs on the energy surfaces in Fig. 1
 (b) illustrate the
 rf-oscillations of the Josephson phase, which in turn, show the curvature of the band
 profile and are detected by the tank circuit coupled to the
 qubit.
The drive frequency $f$ is close to the bare resonance frequency
$f_0 \approx 77$\,MHz of our Nb-tank circuit which has a quality
factor of $Q\approx 370$. The symmetric double loop is coupled -
through the mutual inductance $M$ - to the coil of the double
 spiral tank circuit formed as a planar gradiometer.
The coupling coefficient is $k=M/\sqrt{LL_{\rm T}}\approx 0.4$,
with $M\approx 3.8$ nH, $L\approx 0.7$ nH and the tank circuit
inductance $L_{\rm T}\approx 150$ nH (see Ref. \cite{zang}, where
the behavior of an all-Nb circuit of similar design was
investigated). Due to the coupling to the qubit, the effective
inductance $L_{\rm eff}$ of the resonance circuit is equal to
$L_{\rm T}-M^2L_{\rm J}^{-1}(n,q,\varphi)$.

The Bloch transistor, the  loop and the tank circuit  were
fabricated by electron beam lithography on the same chip. The tank
circuit inductor was fabricated on the basis of Nb technology
\cite{dolata05,zang}, the qubit loop and  the Bloch transistor by
means of the
  two-angle Al shadow evaporation technique.  No special
precautions for the suppression of QP-poisoning of the island -
such as, for example, the  engineering of a barrier-like gap
profile with the island gap value being  greater than the
electrode gap value \cite{chargeexp3,tsai}, or the implementation
of normal-metal QP traps \cite{chargeexp} in the outer electrodes
- were taken. The
 critical currents of the single junctions were approx.
25 nA, with the corresponding value of 45  $\mu$eV for the average
Josephson coupling energy $E_{\rm J0} =(E_{\rm J1}+E_{\rm J2})/2$
($\approx E_{\rm J1}\approx E_{\rm J2}$, whereby the corresponding
values of the single junctions of the transistor  yield the
effective Josephson coupling energy $E_{\rm
J}(\varphi)=\sqrt{E_{\rm J1}^2+E_{\rm J2}^2+2E_{\rm J1}E_{\rm
J2}\cos{\varphi_{\rm dc}}}$). The charging energy $E_{\rm C}$ of
the transistor island has a value of 110 $\mu$eV, so that both
energies $E_{\rm C}$ and $E_{\rm J0}$ are smaller than the value
of the superconducting gap in Al films, $\Delta_{\rm Al}\approx
210$ $\mu$eV.  These values of $E_{\rm J0}$ and $E_{\rm C}$ were
taken from a fitting of the shape of the ground state extracted
from rf-measurements with finite amplitude of the Josephson phase
oscillations, see Ref.\cite{zang} for details. Moreover, the
obtained data agreed well with the corresponding
Ambegaokar-Baratoff and Coulomb-blockade values of similar test
transistors, extracted from dc-measurements of their I-V
characteristics.

Unfortunately,  due to the technical conditions (poor
signal-to-noise ratio for a small amplitude of the phase
oscillations), it was not possible
 to determine the  Josephson energy difference $\Delta
E_{\rm J} = |E_{\rm J1}-E_{\rm J2}| \equiv E_{\rm J}(\varphi_{\rm
dc}=\pi)\ll E_{\rm J0}$. An analysis of the  fitted curves
$\alpha(q)$ shows that their shape for the smallest amplitude
$a=0.28$ of the phase oscillations does not depend on the
asymmetry factor $\Delta E_{\rm J}/E_{\rm J0}$ if this ratio is
smaller than $0.2$. From this we conclude that the asymmetry
factor in our sample is in the range $\Delta E_{\rm J}/E_{\rm
J0}\leq 0.2$.

    In our experiment, which was carried out in a dilution
refrigerator at a base temperature of 20 mK, we measured the phase
angle $\alpha$ between the driving signal $I_{\rm rf}$ and the rf
voltage $V_{\rm rf}$ on the tank circuit. From the
$\alpha$-dependence one can deduce the Josephson-inductance
$L_{\rm J}(q,\varphi)$ by applying the simple formula
\begin{equation}
\tan{\alpha}=k^2Q\frac{L}{L_{\rm J}(q,\varphi)} \label{fa}
\end{equation}
which applies when a drive frequency $f$ is equal to the bare
resonance frequency $f_0$. The measurement of the phase shift,
$\alpha$, for different values of the dc phase and the gate charge
allows the curvature of the energy surface to be mapped.
Figure\,\ref{fig:fig1} (c) shows a periodical dependence of
$\alpha$ both on $\varphi_{\rm dc}=(2\pi/\Phi_0)M I_{\rm dc}$ and
$q=C_{\rm G} V_{\rm G}+q_{\rm offset}$, where we find a
$2e$-periodic gate modulation. As had been expected for the
ground-state shape, a plain gate modulation dependence with a
peak-to-peak amplitude of $0.6$ rad is to be seen in
Fig.\,\ref{fig:fig1} (c) for $\varphi_{\rm dc}=0$. As can be
expected from the theoretical dependence of the ground state
energy, the gate modulation curve for $\varphi_{\rm dc}=\pi$  is
inverted with respect to the $\varphi_{\rm dc}=0$-curve, which is
expressed by the opposite sign of the energy surface curvature. A
closer look at the gate-dependence curve measured at $\varphi_{\rm
dc}=\pi$ reveals, however, a peculiar shape. Before reaching the
zenith of the curve, the phase angle $\alpha$ starts to rise
sharply  and remains in a broad range around $q=e$ at a level that
is even higher than that for $q=0$ (this dependence is also shown
in Fig. 2(a) by symbols).

\begin{figure}
\includegraphics[width=0.89\linewidth]{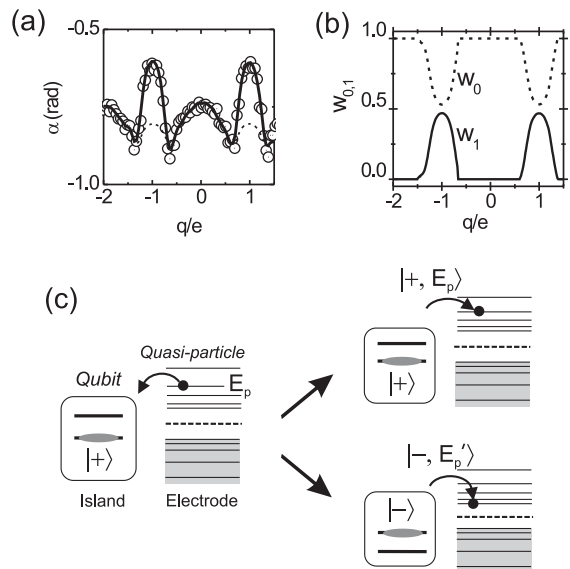}
\caption{\label{fig:fig2}   (a) Comparison between the
experimental (symbols) and the calculated (solid line) gate
modulation dependencies $\alpha(q)$ for an amplitude of
rf-oscillations of $a=0.56$ rad and a phase bias $\varphi_{\rm
dc}=\pi$. (b) Reconstructed occupation probabilities of the ground
state (dotted line) and of the excited  state (solid line),
yielding the modelled dependence in (a). (c) QP-pumping mechanism:
in the left panel,  an unpaired non-equilibrium QP tunnels from
the outer electrode onto the island of the qubit in the ground
state. In the favoured
 process, the QP tunnels - due to the somewhat larger density of
states there - back into a lower-energetic state of the outer
electrode  and, hence, transfers
 energy to the qubit  by exciting it to the upper state (see panel at the lower right). The alternative process, i.e. the tunneling of a QP
back to the outer electrode without the qubit being excited is
shown in the panel at the upper right.}
\end{figure}

 Our theoretical understanding of this behavior bases on the
 assumption of a
statistical mixture of the different quantum states of the qubit.
Here, at the degeneracy point ($q=e$ and $\varphi_{\rm dc}=\pi$),
the states considered are the ground state $A$ and the excited
state $B$, both with even parity of the island, and the ground
state $C$ and the excited state $D$, both with odd parity, as
shown
 by the
 black arcs in  Fig.\,1 (b). In our opinion, the excited odd-parity
state $D$ does not play any significant role, as its excitation
energy is too great (i.e. around $3E_{\rm C}$) and hence much
larger than the energy gap $\Delta E_{\rm J}$ between $A$ and $B$.
We can also rule out a notable contribution of the ground state
$C$ with odd parity, as the admixture of the values of $\alpha(q)$
corresponding to this state  (indicated by the corresponding black
arcs in Fig.\,1 (b))  with small negative curvature cannot yield
the above-mentioned overshooting of the experimental data in the
vicinity of $q=e$. Likewise, we are of the opinion that the
contribution of the odd state is small due to the presumably short
lifetime of a QP in the island, see, e.g.,
Ref.\cite{glazman2,chargeexp3}. Therefore, we modelled the
experimental data as a mixture of the ground state $A$ and the
excited state $B$, with the signs opposite to each other. This was
done by replacing, in Eq.\,(\ref{fa}), the Josephson inductance
with
\begin{equation}
L^{-1}_{\rm J}(q)\rightarrow w_0(q)L^{-1}_{\rm
J}(q)_{|n=0}+w_1(q)L^{-1}_{\rm J}(q)_{|n=1}\label{af}
\end{equation}
(whereby the weight factors are non-negative occupation numbers
which obey the relation $w_{0}(q)+w_{1}(q)=1$) and have  been
fitted subsequently to the measured $\alpha(q)$-dependence, see
Fig.\,\ref{fig:fig2} (a). In this way, we were able to reconstruct
the occupation numbers $w_{0,1}(q)$ of the ground state and of the
excited state, see Fig.\,\ref{fig:fig2} (b). As a result, we found
at $q=\pm e$ an increase in the occupation of the upper state up
to $w_1(e)=0.46$, which remained rather large in a broad range
around this degeneracy point.

 The found steady-state populations $w_{0,1}$ yield the ensemble-averaged values of the Josephson inductance according to Eq.\,(2). The ratio of $w_1$ to $w_0$ reflects the competition
between the rates of the QP tunneling and the energy relaxation
 of the qubit. Such a relaxation also occurs due to the coupling to the environmental degrees of freedom
 (e.g. flux and gate control lines, external magnetic field, background charge, etc.).
 From our measurements at $\varphi_{\rm dc}=\pi$ we can conclude that the relaxation due to the environmental degrees of freedom which reduces $w_1$ and increases $w_0$ is not dominant, because in our case $w_0\approx w_1\approx 0.5$.
 Since  we measure with our technique only
averaged values of the phase
 shift $\alpha$, we are not able  to find absolute values of the relaxation rates. Beyond this, an accurate
determination of the different relaxation rates is only possible
when the distribution of the non-equilibrium QPs is known with
regard to their energy. To explain this mixture effect, we rule
out a thermal activation
 of the excited state at the given base temperature, which is below 100 mK, because at
 such low
 temperatures   such high $w_1$-values can never be achieved, i.e. the Boltzmann-factor for the excited state is negligible.
 Besides, the broad  range of  the gate charge where mixture occurs (more than 40\% of the gate  charge
 period)
 cannot be achieved by such a weak thermal excitation. This is shown by the following estimation:
 for a finite detuning  from the degeneracy point $q=e$,
 the gap energy increases roughly like $\Delta E=4E_{\rm
 C}|q/e-1|$ and the factor $\exp{(-\Delta E/k_{\rm B}T)}$ is therefore strongly
 suppressed at a finite detuning $\delta q\equiv q/e-1=0.1$ for   $E_{\rm
 C}\approx 110$  $\mu$eV and $T=100$ mK.

 Another possible
 explanation for the observed mixture effect could be Landau-Zener (LZ) tunneling due to the
 rf-drive which leads to a periodic passing of the degeneracy point
 (at $\varphi_{\rm dc}=\pi$ and $q=e$).  When we take the broad  range of the
mixture effect into account, LZ tunneling seems to be unlikely,
since it is exponentially suppressed for a finite $\delta q$. When
we apply the parameters of our experiment, the estimated LZ
tunneling probability  $p_{\rm LZ}$ is negligible already for
small values of $\delta q=0.1$:
\begin{eqnarray}
p_{\rm LZ} & = & \exp{\left[ -\pi\frac{\Delta E^2}{2\hbar
\dot{E}}\right]}=\exp{\left[-\frac{\Delta E^2}{4a\hbar f E_{\rm
J0}}\right]}\ll 1
 \end{eqnarray}
 for $E_{\rm J0}\approx 45$ $\mu$eV,  $E_{\rm
 C}\approx 110$  $\mu$eV and an intermediate amplitude $a=0.5$ of
 rf-oscillations. On the other hand,  experimentally we still observe   a
strong contribution of the excited band at $\delta q=0.1$.

 A further argument against LZ tunneling is the behavior of the curves
 measured at  $\varphi_{\rm dc} = 0$ and at values of $q$ around $e$ (except for those at $q=e$, which will be discussed later). They clearly
 show the effect of qubit excitation, whereas the qubit energy level spacing is, in this case, much larger
 than in the case of  $\varphi_{\rm dc} =\pi$.

 Following a model given by Ref.\cite{glazman2}, we relate the
 observed
mixed state  to a strong coupling of non-equilibrium QPs to the
qubit system, which allows the transfer of energy to the qubit and
thus an excitation of its upper state. This QP pumping of the
qubit can be considered as a cycle in which an unpaired
non-equilibrium QP tunnels onto the island while the qubit is in
the ground state $|+\rangle$, see Fig.\,\ref{fig:fig2} (c). As
soon as the QP enters the transistor island, it changes the excess
charge and induces in this way an instantaneous change of the
working point and of the energy level splitting. Only when the QP
tunnels to a lower-energetic state of the electrode, a transfer of
energy to the qubit system occurs, exciting it to the upper state
$|-\rangle$, whereas a QP  tunneling back to the initial state of
the electrode does not induce any  excitation of the qubit.
According to Ref. \cite{glazman2}, the former process should
prevail due to the greater density of the states that are close to
the edge of the QP band in the energy spectrum
\cite{comment_pumping}. Of course, such  non-equilibrium QPs
should have an energy which is at least by the value of $E_{\rm
J}$ larger than the value of the superconductor energy gap, in
order to transfer energy to the qubit. This would mean, however,
that a QP having a lower energy could enter the island as well,
but in that case the QP should leave the island without exciting
the qubit. Trapping the QP in the island is only possible if the
superconductor energy gap of the island is lower than that of the
electrodes. Due to particularities of our sample fabrication, the
energy gaps of the electrodes and of the transistor island are
almost equal. Moreover,we presume that the gap of the island is
even slightly
 larger. We assume that QPs having an energy
lower than $\approx E_{\rm J}$ above the energy gap are available
in the outer electrodes and that their relaxation to the gap edge
occurs both via interaction with the lattice of the electrodes and
via the traveling onto the island and back into the electrodes
with simultaneous excitation of the qubit.

 The pumping process depends strongly on the qubit bias and on
the amplitude of the rf-oscillations.
\begin{figure}
\includegraphics[width=0.85\linewidth]{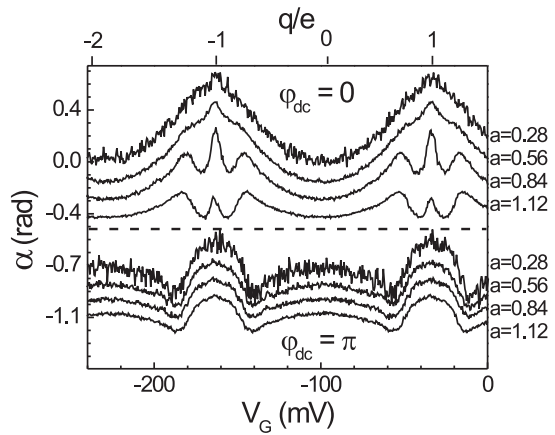}
\caption{\label{fig:fig3} Periodic gate modulation curves
$\alpha(V_{\rm G})$ for different values of the amplitude of the
rf-oscillations, measured at two values of the dc phase bias. For
the purpose of clarity, the curves are vertically shifted. The
upper horizontal axis presents the values of the island charge.}
\end{figure}
 This is demonstrated in the  gate modulation curves $\alpha(V_{\rm G})$ in Fig.\,3 for
different values of the dc phase bias. Interestingly, when
increasing  the amplitude of the rf-oscillations for $\varphi_{\rm
dc}=0$ (top part of the plot) from 0.28 rad to 1.12 rad, two dips
appear around the sharp peak at $q=e$. This is in contrast to the
behavior  we have described for $\varphi_{\rm dc}=\pi$ (bottom
part of the plot). Here, the gate modulation curves practically do
not depend on the amplitude $a$ of the flux oscillations.  We
interpret these observations in the following way: for
$\varphi_{\rm dc}=0$, the pumping process
  is not active for small rf-oscillations. It
starts to develop at increasing values of $a$, and the shape of
the curves shows a strong dependence of this mechanism on the
charge $q$.  In particular, the pumping is partially suppressed at
$q=e$ even for rather large values of $a$. Therefore, we  extend
the theory of Ref.\cite{glazman2} by establishing a selection rule
for the QP transitions between the ground state and the excited
state for certain flux-bias-values.

\begin{figure}
\includegraphics[width=0.69\linewidth]{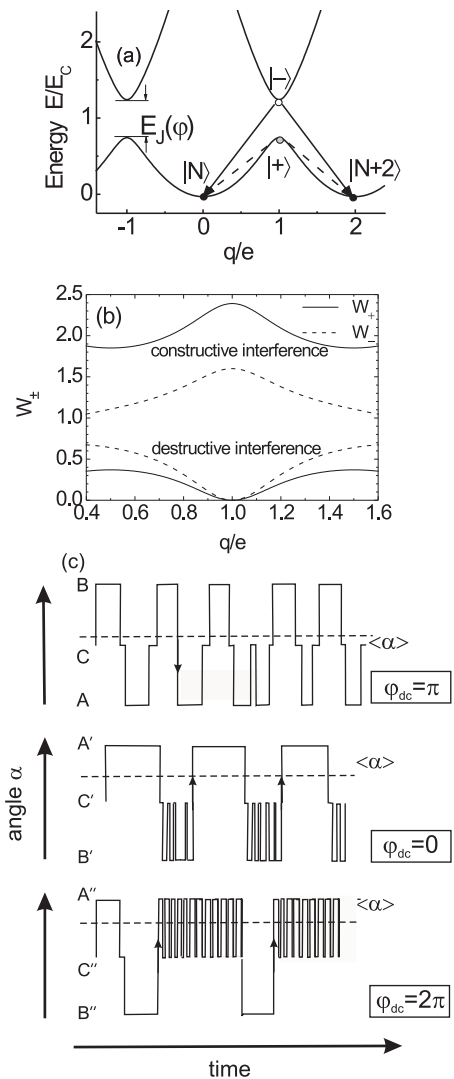}
\caption{\label{fig:fig4} (a) Energy  diagram of the qubit,
displaying the interfering paths of the excess QPs  tunneling
either  - as a quasi-electron (to the left) - or  - as a
quasi-hole (to the right). (b) The rate factors for the intraband
transition probability $W_+$ and interband transition probability
$W_-$, calculated for the experimental parameters of the circuit
for the integer number $m$ of flux quanta in the loop as a
function of the gate charge. For even $m$, the destructive
(constructive) interference takes place for the intraband
(interband) transitions, and for odd $m$ vice versa. (c) Schematic
time-traces of the cyclic transitions between qubit states at
$q=e$ for different values of $\varphi_{\rm dc}$ shown in Fig.\,1
(b). The thin arrows indicate the qubit relaxation  in contrast to
the quasiparticle-induced transitions.}
\end{figure}

The QP transitions are described by the tunneling Hamiltonian,
$H_T$, which consists of two Hermitian conjugate terms and is
considered to be a perturbation term, see  Ref. \cite{glazman2}.
The Josephson coupling term describing the tunneling of Cooper
pairs is naturally included in the Hamiltonian of the qubit. To
describe this peculiar behavior observed in the experiment, we
should take  the interference effect into account which occurs
when QPs tunnel onto the island. Let us assume that the energies
of the non-equilibrium QP in the electrode $E_p =
(\Delta^2+\xi_p^2)^{1/2}$ and in the island $E_k =
(\Delta^2+\xi_k^2)^{1/2}$ exceed  the gap value $\Delta$ only
slightly, i.e. the corresponding electron energies are
sufficiently small, $|\xi_p|, |\xi_k| \ll \Delta$. In this case
the absolute values of the Bogoliubov-Valatin coherence factors of
the corresponding single QP states \cite{Schrieffer} are almost
equal to each other,
\begin{equation}
u_p^2 \approx v_p^2 \approx u_k^2 \approx v_k^2 \approx 0.5.
\label{u-v}
\end{equation}
This means that  QPs can tunnel onto the island both as an
electron-like particle and as a hole-like particle. The qubit
biased by the gate charge $q = (N+1)e$ with $ N=0, \pm 2,...$
switches into the odd (ground) state $|o\rangle$ following two
paths, $|\pm\rangle \rightarrow |N\rangle$ and $|\pm\rangle
\rightarrow |N+2\rangle$, as shown in Fig.\,\ref{fig:fig4} (a) by
arrows. The sign  "$+$" ("$-$") stands for the even ground state
(excited state) of the qubit. These states represent coherent
superpositions of states with a different number of Cooper pairs
on the island (see, e.g., Eq.\,(5) in Ref.\cite{Zorin-JETP}).
These  phase factors in the series do not only depend on the
particular state of the qubit, but also on the value of the
magnetic flux $\Phi$ which penetrates the qubit loop and thus
determines  the overall Josephson phase
$\varphi=\varphi_1+\varphi_2$. This dependence is expressed most
simply in the case of the small ratio $E_J/E_c$, i.e.,
\begin{equation}
|\pm\rangle = (|N\rangle \pm e^{i \varphi/2}|N+2\rangle
)/\sqrt{2}, \label{+-}
\end{equation}
while in the general case a low-weight admixture of higher order
terms with corresponding signs and phases, i.e., $(\pm1)^k e^{i k
\varphi/2}|N+2k\rangle$, where integer $k=-1,-2,-3,\dots$ and
$k=2,3,4,\dots$, should also be added. One can see that in
contrast to the case of a plain Cooper pair box which, in the
degeneracy point, has a symmetric (antisymmetric) wave function in
the ground (excited) state, the symmetry property of the states
$|\pm\rangle$ in the box of the SQUID configuration alternates
with the flux value $\Phi$. Especially, for  values which
correspond to an even number of flux quanta, $\Phi=m \Phi_0, m=0,
\pm 2,...$, yielding $e^{i \varphi/2}=1$, the ground state
(excited state) of the qubit is described by a symmetric
(antisymmetric) wave function, whereas for odd numbers of flux
quanta, $m=\pm 1, \pm 3...$, the ground state (excited state) is
antisymmetric (symmetric) \cite{Averin-comment}. Therefore, at any
integer value of flux quanta in the loop, at least  one of the
states (ground state or excited state) is necessarily symmetric at
$q=e$. As we  see below, this property plays an important role for
establishing the selection rule for the rate of the QP tunneling
which includes these states.

The  operators  causing transitions in the qubit are the charge
shift operators $e^{\pm i\phi/2}$, where
$\phi=(\varphi_1-\varphi_2)/2$ is the variable conjugate to the
island charge operator (see, e.g., Ref.\cite{Ingold}). These terms
appear in the tunneling Hamiltonian terms as phase factors of the
creation- and annihilation-operators  associated with the island.
Physically, these phase factors ensure a coupling between the
tunneling particle and the environment, i.e. the charge degree of
freedom of the qubit, cf. Ref.\cite{Ingold}. Applying the Fermi
Golden rule for calculating the transition rates of the QP
tunneling onto the island, we obtain
\begin{equation}
\Gamma_\pm^{\rm{in}}(E_p,E_k) = 2\pi |\langle E_p,\pm|H_T|o,E_k
\rangle|^{2} \delta(E_p+E_{\pm}-E_k). \label{gamma-rate}
\end{equation}
Taking into account the  simplifying assumption Eq.\,(\ref{u-v}),
the matrix elements in Eq.\,(\ref{gamma-rate}) can be presented as
\begin{eqnarray}
 && |\langle E_p,\pm|H_T|o,E_k \rangle|^{2}
\nonumber\\
 &&\qquad \qquad  = |t_{pk}|^2 \,[ \,|e_{\pm}|^2|u_p|^2|u_k|^2+|e_{\pm}'|^2|v_p|^2|v_k|^2
\nonumber\\
 &&\qquad  \qquad - e_{\pm}e_{\pm}'u_pu_kv_pv_k
 - e_{\pm}'e_{\pm}u_pu_kv_pv_k ]
 \nonumber\\
 &&\qquad  \qquad = |t|^2 \,|e_{\pm}-e_{\pm}'|^2/4 , \label{H_T-matrix}
\end{eqnarray}
where the matrix elements are $e_\pm = \langle
\pm|e^{i\phi/2}|N\rangle$ and $e_\pm'
=\langle\pm|e^{-i\phi/2}|N+2\rangle$. Here $|t_{pk}|^2 \equiv
|t|^2$ is the square of the absolute value of the tunnel matrix
element which is only  weakly dependent on the energies $E_p$ and
$E_k$. Therefore, the net rate of the QPs tunneling onto the
island can be expressed as
 $\Gamma_\pm^{\rm{in}}\propto (\delta_{\rm r}/R_{\rm T})W_\pm$,
 with $W_\pm=|e_\pm - e_\pm'|^2$ (see the plots in Fig.\,\ref{fig:fig4} (b))
 and $\delta_{\rm r}$ being
 the QP level spacing in the island.

Since at $q=e$ one of the qubit states $|+\rangle$ (for even $m$)
or $|-\rangle$ (for odd $m$) is described by a symmetric wave
function, the matrix elements $e_+$ and $e_+'$ (for even $m$) or
$e_-$ and $e_-'$ (for odd $m$) are identical, which leads to zero
 value for $W_+$ or $W_-$. This manifests itself in the zero
minimums in the two lower curves in Fig.\,\ref{fig:fig4} (b), as
well as in the suppression of the corresponding transition rates
$\Gamma_+^{\rm{in}}$ or $\Gamma_-^{\rm{in}}$, respectively. In
contrast to this, the transitions in which asymmetric qubit states
are involved have an increased probability (the two upper curves
in Fig.\,\ref{fig:fig4} (b)) due to the constructive interference
which results from the opposite signs of the corresponding matrix
elements.
The effect of the complete destructive interference  occuring in
transitions involving the symmetric states of the qubit is
formulated as a selection rule. Note that the same selection rule
is applied to transitions which are associated with the tunneling
of the QP from the island back into the electrode with
simultaneous switching of the qubit into a state with symmetric
wave function. If finite QP energies $\xi_q$ and $\xi_k$ above the
superconductor gap are taken into account, this leads to a
somewhat unbalanced ratio of the coherence factors $u$ and $v$ of
Eq.\,(4) and, hence, to an incomplete destructive interference by
which the selection rule is violated.

The  destructive interference described is supposed to have a
suppressing effect on the cyclic qubit excitation. In our scenario
of the qubit pumping at $\varphi_{\rm dc}=\pi$, the rates of both
intraband and interband transitions were assumed to be much larger
than the rate $\gamma_r$ of the qubit relaxation $B\rightarrow A$.
In this case the pumping cycle is closed and the two qubit states
$A$ and $B$ are almost equally populated at $q=e$, see Fig.\,2
(b). In this case, the average phase angle $\alpha$ is positioned
close to, but slightly higher than the level of the state $C$,
manifesting itself in the experimental $\alpha-V_{\rm G}$-curve in
Fig.\,2 (a) as an overshooting around $q=e$. This process is
illustrated by the time-trace of repeated qubit transitions in the
bottom panel of Fig.\,4 (c) for $\varphi_{\rm dc}=\pi$, i.e. for
the case where neither of the states is symmetric. Let us consider
now the phase bias $\varphi_{\rm dc}=0$, where - as has been
discussed before - the qubit ground-state has a symmetric wave
function with a corresponding low rate $\gamma_{\rm symm}$ for
transitions into this state. Due to the effect of the destructive
interference  (denoted by the dashed arrows in Fig.\,1 (b)and
Fig.\,4 (a)), the rate of intraband transitions
$A'\rightleftarrows C'$ is presumably much slower than the qubit
relaxation. As a result, the qubit remains in the even symmetric
state $A'$ for a sufficiently long time
 or switches fast between the even antisymmetric
state $B'$ and the odd state $C'$ on an average time scale
$\gamma_r^{-1}$ (see the time-trace for $\varphi_{\rm dc}=0$ in
the central panel of Fig.\,4 (c)). If, on the other hand, the flux
bias is odd, e.g. $\varphi_{\rm dc}=2\pi$, the excited state of
the qubit is symmetric, which leads to a strong reduction in the
interband transitions $B''\rightleftarrows C''$. The schematic
time-trace in the central panel in Fig.\,4 (c) shows for this case
a long series of rapid transitions $A''\rightleftarrows C''$,
interrupted by rare (with a rate of $\sim\gamma_{\rm symm}$)
transitions into the state $B''$. The escape from the excited
state $B''$ with an average occupation time $\gamma_r^{-1}$ mainly
occurs via qubit relaxation $B''\rightarrow A''$. When we
summarize the case of an integer dc phase bias $\varphi_{\rm
dc}=2\pi m$, the sufficiently high rate of relaxation with respect
to the small rate due to the destructive interference
($\gamma_r\gg \gamma_{\rm symm}$) at $q=e$ ensures that
$\langle\alpha\rangle$ is close to the value corresponding to the
ground state $A'$ or $A''$. Hence, this average value of the phase
angle shows that the destructive interference effect suppresses
the pumping cycle.

This  interference effect can qualitatively explain the shape of
  the $\alpha-V_g$ dependencies measured at zero
flux value and shown in the upper part of Fig.\,\ref{fig:fig3}.
The two upper curves, measured at a small amplitude $a$ of
rf-oscillations, exhibit a shape which corresponds almost entirely
to the expected form of the ground state. This behavior is
related,  first of all, to relatively small deviations of the
Josephson $\varphi$ from the zero value, in the vicinity of which
the destructive interference is  strongest and, secondly, to a
relatively large QP energy ($\sim 2E_{\textrm{J0}}$) which is
required for the excitation of the qubit. The increase in the
amplitude $a$ leads to distinct deviations of the Josephson phase
 $\varphi$ from the zero value, so the destructive interference is
still efficient only in a narrow region around the value $q=e$.
This effect manifests  itself in a sharp peak centered around the
value $q=e$, whereas slight deviations from this value lead to the
two side dips and are an indication of a significant admixture of
the excited state. The described behavior can be understood as a
significant weakening of the destructive part of the interference,
which is due to the deviation from the optimum point both in the
dc phase and charge. Moreover, the excitation qubit energy is in
this case  somewhat lower than $\sim 2E_{\textrm{J0}}$. Note that
in the case of a semi-integer flux quantum bias (the respective
curves are presented in the bottom part of Fig.\,\ref{fig:fig3}
for $\varphi_{\rm dc}=\pi$), the effect of destructive
interference does not show. This is understandable, because
neither of the wave functions describing the ground state and the
excited state possesses the property of symmetry (see Eq.\,(5) in
which the parameter is $\varphi=\pi$).

In conclusion, we found the effect of  mixing of the qubit
 states, which we explained by an energy transfer between the
 non-equilibrium quasiparticles and the two-level qubit
 system. To explain the qubit excitation observed in the experiment, we  applied a picture of the stochastic  tunneling  of
 single quasiparticles  and derived
 a selection rule for the   quasiparticle tunneling at the magic points $q=e$ of
 the qubit. This selection rule is applicable to transitions in which qubit states with a symmetric wave function are involved, and it
 strongly suppresses the mechanism
 of the qubit pumping. Our set-up only permitted a continuous readout
 of the - therefore - averaged qubit state. A pulsed readout set-up with a higher drive frequency
 could help to discriminate the single tunneling events and allow probing
  the ground state and the excited state of the qubit separately.
\begin{acknowledgments}
We are indebted to D.\,V.~Averin for his enlightening comments on
this work. We wish to thank R. Lutchyn and L. Glazman for their
useful comments on the selection rule for the quasiparticle
transitions and also M.~Wulf and J.~Niemeyer for discussions on
the results of the experiment. We also acknowledge S.\,V.~Lotkhov
for the test sample characterization, H.-P. Duda and R. Harke for
valuable technical assistance,  B. Egeling and R. Wendisch for
their support in the PECVD and CMP processes as well as
Th.~Weimann and P.~Hinze for their support with the e-beam
writing. This work was supported by the European Union (project
EuroSQIP).
\end{acknowledgments}


\begin{thebibliography}{00}
\bibitem{Averin-Likharev} D.\,V.~Averin and K.\,K.~Likharev, in \it Mesoscopic Phenomena in
Solids, \rm edited by B.\,L.~Altshuler, P.\,A.~Lee, and
R.\,A.~Webb (Elsevier, Amsterdam, 1991), p.175.

\bibitem{tuominen} M.\,T.~Tuominen, \emph{et al.},
Phys. Rev. Lett. {\bf 69}, 1997 (1992).

\bibitem{chargeexp} P.~Joyez, \emph{et al.}, Phys.~Rev.~Lett. {\bf 72}, 2458 (1994).

\bibitem{CP-R-electr-theo} A.\,B.~Zorin, Phys. Rev. Lett. {\bf 76}, 4408 (1996).

\bibitem{CP-R-electr-exp-theo} A.\,B.~Zorin,  \emph{et al.}, J. Supercond. {\bf 12}, 747
(1999); A.~B.~Zorin, Phys. Rev. Lett. {\bf 86}, 3388 (2001).

\bibitem{CP-electr-HUT} M.\,A.~Sillanp\"a\"a, L.~Roschier and P.~Hakonen,
Phys. Rev. Lett. {\bf 93}, 066805 (2004).

\bibitem{Geerligs} L.\,J.~Geerligs,  \emph{et al.},
Z.~Phys. B: Condens. Matter {\bf 85}, 349 (1991).




\bibitem{Amar} A.~Amar \emph{et al.}, Phys. Rev.
Lett. {\bf 72}, 3234 (1994).

\bibitem{Bouchiat} V.~Bouchiat \emph{et al.}, Phys. Scripta {\bf T76}, 165
(1998).

\bibitem{lehnert} K.\,W.~Lehnert \emph{et al.}, Phys.~Rev.~Lett. {\bf 91},
106801 (2003).

\bibitem{lukens} J.~M\"annik and J.\,E. Lukens, Phys.~Rev.~Lett. {\bf 92}, 057004 (2004).

\bibitem{echternach} A.~Guillaume \emph{et al.}, Phys. Rev. B {\bf 69}, 132504 (2004).

\bibitem{aumentado} O.~Naaman and J.~Aumentado, Phys. Rev. B {\bf 73}, 172504 (2006).

\bibitem{qubit-NEC}
Y.~Nakamura, Yu.\,A.~Pashkin and J.\,S.~Tsai, Nature {\bf 398},
768 (1999).

\bibitem{qubit-Saclay}
D.~Vion,  \emph{et al.}, Science {\bf 296}, 886 (2002).

\bibitem{qubit-Chalmers} T.~Duty \emph{et al.},
Phys. Rev. B {\bf 69}, 140503(R) (2004).

\bibitem{Zorin-JETP} A.\,B.~Zorin, JETP {\bf 98}, 1250 (2004).


\bibitem{cpqb2} A.\,B.~Zorin, Physica C {\bf 368}, 284 (2002).

\bibitem{imt2} D.~Born \emph{et al.}, Phys. Rev. B
{\bf 70}, 180501(R) (2004).

\bibitem{zang} H.~Zangerle \emph{et al.}, Phys. Rev. B {\bf 73}, 224527 (2006).

\bibitem{glazman2} R.\,M.~Lutchyn, L.\,I.~Glazman, and A.\,I.~Larkin, Phys. Rev. B {\bf
74}, 064515 (2006).




\bibitem{sqc} K.\,K.~Likharev and A.\,B.~Zorin, J. Low Temp. Phys.
{\bf 59}, 697 (1985); D.\,V.~Averin, A.\,B.~Zorin, and
K.~\,K.~Likharev, Sov. Phys. JETP {\bf 88}, 697 (1985).


\bibitem{dolata05}  R.~Dolata \emph{et al.}, J. Appl. Phys. {\bf 97}, 054501 (2005).


\bibitem{chargeexp3} J.~Aumentado \emph{et al.}, Phys. Rev. Lett. {\bf 92}, 066802 (2004).

\bibitem{tsai} T.~Yamamoto \emph{et al.}, Appl. Phys. Lett. {\bf 88}, 212509 (2006).

\bibitem{comment_pumping} A finite rate of the qubit relaxation  works
opposite to the pumping direction.

\bibitem{Schrieffer} J.\,R.~Schrieffer, \emph{Theory of
Superconductivity} (Perseus, Oxford,1999).

\bibitem{Averin-comment} This property of the SQUID-configuration qubit was
pointed out to the authors by D.\,V.~Averin.


\bibitem{Ingold} G.\,L.~Ingold and Yu.\,V.~Nazarov, in \it Single Charge Tunneling, \rm edited by H.~Grabert,  and
M.H. Devoret (Plenum Press, New York, 1992), p. 21.



\bibitem{matrix-elem-e} Technically, these matrix elements were computed
using the Fourier-series expansions of the Bloch-wave eigenstates
of the qubit Hamiltonian, which were derived in Ref.\cite{sqc}.

\end{thebibliography}

\end{document}